# Engineering the Temporal Dynamics with *'Fast'* and *'Slow'* Materials for All-Optical Switching

Soham Saha[1], Benjamin Diroll[3], Mustafa Goksu Ozlu[1], Sarah N. Chowdhury[1], Samuel Peana[1], Zhaxylyk Kudyshev[1], Richard Schaller[3], Zubin Jacob[1,2], Vladimir M. Shalaev[1,2], Alexander V. Kildishev[1,2], and Alexandra Boltasseva[1,2]*

[1] *School of Electrical and Computer Engineering, Birck Nanotechnology Center, Purdue University, West Lafayette, IN, USA*
[2] *Purdue Quantum Science and Engineering Institute, Purdue University, West Lafayette, IN, USA*
[3] *Argonne National Laboratory, Lemont, IL 60439, USA*
*aeb@purdue.edu

## 1. Abstract

All-optical switches offer advanced control over the amplitude, phase, and/or polarization of light at ultrafast timescales using optical pulses as both the carrying signal and the control. Limited only by material response times, these switches can operate at terahertz speeds which is essential for technology-driven applications, such as all-optical signal processing and ultrafast imaging, as well as for fundamental studies, such as frequency translation and novel optical media concepts such as photonic time crystals. In conventional systems, the switching time is determined by the relaxation response of a single active material, which is generally challenging to adjust dynamically. This work demonstrates that the zero-to-zero response time of an all-optical switch can instead be varied through the combination of so-called "fast" and "slow" materials in a single device. When probed in the epsilon-near-zero (ENZ) operational regime of a material with a slow response time, namely, plasmonic titanium nitride, the proposed hybrid switch exhibits a relatively slow, nanosecond response time. The response time then decreases as the probe wavelength increases reaching the picosecond time scale when the hybrid device is probed in the ENZ regime of the faster material, namely, aluminum-doped zinc oxide. Overall, the response time of the switch is shown to vary by two orders of magnitude in a single device and can be selectively controlled through the interaction of the probe signal with the constituent materials. The ability to adjust the switching speed by controlling the light-matter interactions in a multi-material structure provides an additional degree of freedom in the design of all-optical switches. Moreover, the proposed approach utilizes "slower" materials that are very robust and allow to enhance the field intensities while "faster" materials ensure an ultrafast dynamic response. The proposed control of the switching time could lead to new functionalities and performance metrics within key applications in multiband transmission, optical computing, and nonlinear optics.

## 2. Introduction

Switchable optical devices enable real-time control over the polarization, amplitude, or phase of light[1–6]. In all-optical switching, a light pulse interacts with the device's constituent material, changing its optical response. For example, an intraband pump can energize free electrons in a transparent conducting oxide (TCO), increasing their effective mass, making the materials less absorptive[7–11] and thus increasing overall transmission. Conversely, an interband pump can generate photocarriers that increase the metallicity and absorption of TCOs[12,13]. This mechanism is often suggested for dynamic photonic devices capable of modulating the amplitude or phase of light passing through it. Importantly, all-optical switches can operate without the resistive-capacitive delays of electronic circuits. Instead, the speed of optically-induced permittivity modulation is limited by the relaxation mechanisms of the switching material, which can range from a few femtoseconds[7,14,15] to several nanoseconds[16,17]. The fast permittivity change enables several interesting phenomena and applications that are not achievable by other means of switching. Large and rapid modulation of the dielectric permittivity in TCOs has led to nonreciprocal optical devices[18], photon acceleration[19], and ultrafast optical switching[20,21], to name a few examples.

Recently, field-enhanced slow-light effects in epsilon-near-zero materials, including TCOs, have further enhanced these nonlinear optical phenomena by several orders of magnitude[22,23]. The wavelength range where the real part of the material dielectric permittivity changes its sign is known as the Epsilon-Near-Zero regime. Strong light-matter interaction due to the high Purcell effect and the slow group velocity of light near the ENZ point enables a plethora of interesting optical phenomena without the need for complex composite structures. Examples of such phenomena include dramatic reflectance and transmittance modulation[12,24], remarkably strong nonlinearity enhancement[7,25–27], time refraction[28], broadband[29] and narrowband[30] absorption, optical time reversal[31] and high-harmonic generation[32], in addition to on-chip modulators[33,34]. When an ENZ medium is fabricated on a reflective metal, free-space light can couple into radiative bulk plasmon-polaritons, called Ferrell-Berreman (FB) modes[35,36]. Strong light-matter interaction in such multilayer cavities has been employed to demonstrate broadband absorbers[37] and polarization switches[38].

Any optical switching device relies on two major functionalities: the magnitude of the dielectric permittivity change, which governs the modulation depth, and the overall material response time that governs the switching speed. There have been numerous studies on modulating the permittivity of materials in a variety of ways, including electrical[34,39–42], thermal[43–46], and optical methods[8,9,13,47–51]. These studies have focused on the scaling of modulation with power[25,41,52,53], enhancement with engineered structures[7,13], and modulation limits[44,54,55].

Thus far, few studies have focused on engineering the switching speed of tunable devices, which is generally fixed and/or defined by the response times of the material constituents. The challenge in controlling the speed of all-optical switches is that the relaxation time of the switching material is an intrinsic property. As a rule, this property can be changed or adjusted at the film growth/fabrication step but not dynamically during device operation. Thus, the overall switching response is assumed to be fixed after the device is fabricated. In one such study, the relaxation time of cadmium oxide-based switches was decreased by increasing the doping concentration[10]. Epsilon-near-zero switches utilizing aluminum-doped zinc oxide have a picosecond response[12] when controlled by an interband pump, but a femtosecond response[14] when controlled by an intraband pump. All-optical switches with higher quality factors generally have a longer response time than those with a lower quality factor[56]. The carrier dynamics of gold nanorods can be somewhat modified by pumping them at varying wavelengths or angles, varying the switching time to a limited degree[57]. Overall, the range of switching times obtained in these prior studies are within the same timescale order, determined by the carrier dynamics in the single switching material.

Interestingly, several works studying the carrier dynamics of different materials report response times that vary with probe wavelength. For example, in pump-probe studies of gold, longer wavelength probes exhibit faster relaxation times as they interact with states that relax faster than those probed by shorter wavelengths[58]. In black phosphorus, for the same pump wavelength, the decay time of sub-bandgap probes decreases at shorter wavelengths[59]. In titanium nitride, a sub-picosecond electron-phonon response is only observed when the probe wavelength is close to the epsilon-near-zero (ENZ) wavelength[60]. These experiments highlight that the chosen probe, or device operational wavelength, has an important role in the observed dynamics of a structure.

Extending this idea, we demonstrate a novel, device-level approach to controlling all-optical switching speed in a dual-active-material structure employing the Epsilon Near Zero properties of the materials.

This work utilizes two technologically relevant, robust materials whose relaxation dynamics have been studied in great detail, namely plasmonic titanium nitride (TiN) and aluminum-doped zinc oxide (AZO). TiN has an overall relaxation time spanning nanoseconds[61,62], while AZO has a much faster relaxation time spanning picoseconds[12,14]. Employing these two materials, we develop a double-resonant device that supports a radiative ENZ mode in the TiN layer and a Ferrell Berreman mode in the AZO layer [35,63].

A 325-nm-wavelength optical pump simultaneously excites electrons in the TiN and AZO, modulating their permittivity, and resulting in transient reflectance modulation. The magnitude of the reflectance modulation is highest near the epsilon-near-zero regions of the corresponding materials. Proximal to the ENZ wavelength of TiN, the device switching time is nanosecond-speed, following the carrier dynamics of TiN. Close to the ENZ of AZO, the switching time is picoseconds, as expected from AZO dynamics. The overall zero-to-zero response decreases with increasing wavelength – as the probe light increasingly interacts with the faster material (AZO). These results illuminate the important role of light-matter interaction with different materials in an optically modulated device. This work also develops a comprehensive understanding of the temporal dynamics of devices with multiple functional materials, paving the way for a wide array of advanced time-varying metasurface applications.

The following section reports on the experimental findings. After this, the paper is structured to highlight the transition from a steady-state planar multi-material device to a dynamic one. First, we describe the design of the planar device and its steady-state characterization. Then, we discuss the pre-characterization of the titanium nitride and aluminum-doped zinc oxide films individually via pump-probe spectroscopy. We investigate their switching speeds and extract their relaxation constants. This study is followed by the dynamic characterization of the multi-material all-optical switching device. By combining the results of these studies, we developed a model for the effective temporal response of the all-optical switch. We incorporated the contributions of the individual constituent materials to the overall device switching dynamics. Finally, we discuss the experimental findings and their implications for designing more complex all-optical switches.

## 3. Results

We developed a TiN-AZO FB resonator by depositing a 130-nm-thick layer of TiN on a silicon substrate, followed by the growth of a 250-nm-thick AZO layer (**Fig.1a**). Supporting Information Sections 1-3 elaborate on the growth parameters and the optical characterization of the fabricated films. For our pump-probe experiments, we use a normal-incidence 325-nm-wavelength pump beam, which is absorbed in the AZO and the TiN layers. This effect is evident from the absorbance plots obtained from finite-element frequency domain (FEFD) simulations (COMSOL Multiphysics) (**Fig. 1b**, left). The probe light is shone at a 50º angle of incidence. At visible wavelengths, the probe mostly interacts with the TiN, whereas at near-infrared wavelengths, it interacts strongly with the AZO (**Fig. 1b**, right). The structure shows two reflectance dips for *p*-polarized light near the ENZ points of TiN and AZO, evident in **Fig. 1c**.

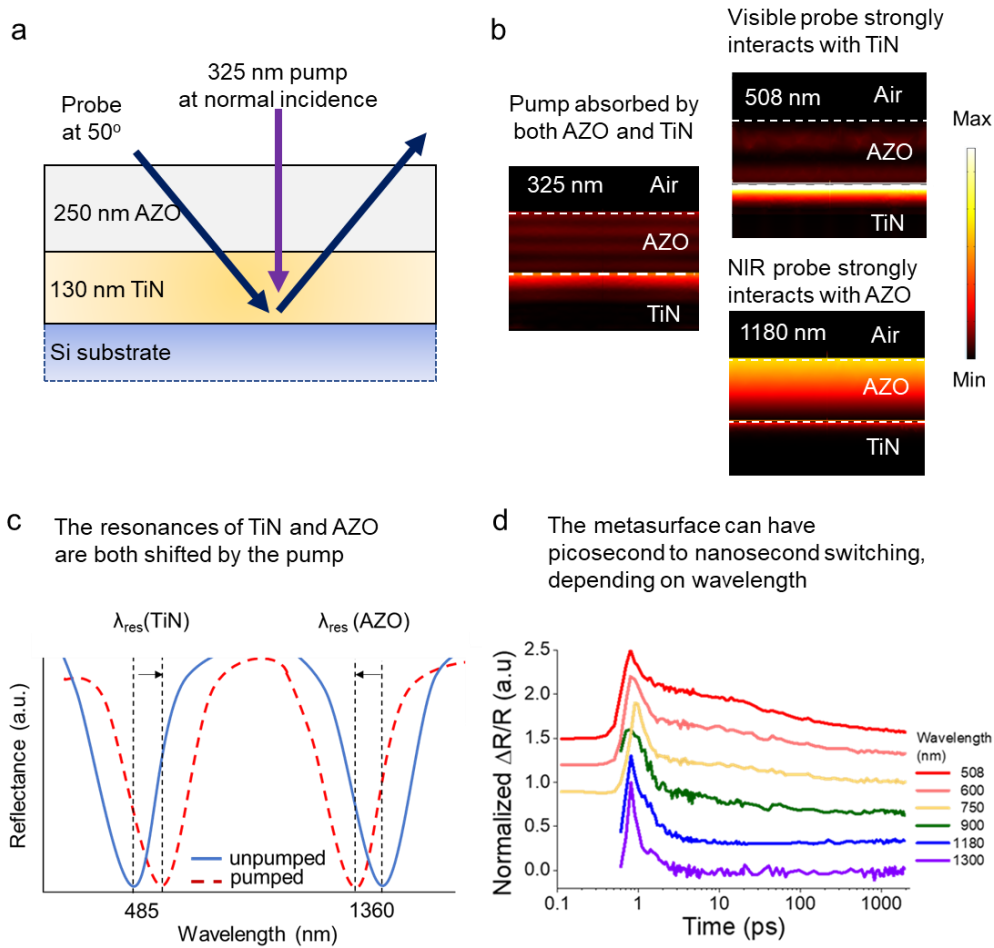

Figure 1 (a) The double-resonant two-layer device comprises a 130-nm-thick TiN layer grown on silicon, with a 250-nm-thick AZO layer deposited on top. (b) At normal incidence, the 325-nm wavelength pump is strongly absorbed in the AZO and TiN by exciting electrons in both materials. The materials interact most strongly with light near their respective ENZ wavelengths. Thus, at visible wavelengths, most of the probe interacts with the TiN, whereas the NIR probes interact more with the AZO layer. (c) The pump causes the reflectance spectrum to redshift at visible wavelengths, whereas, at near-infrared wavelengths, the pump blueshifts the reflectance spectrum. (d) The mechanism of fast and slow switching: TiN has a nanosecond response time, and AZO a picosecond response time. When excited by the same pump, the device has a slower observed response time in the visible probe wavelengths, where its behaviour is dominated by the TiN response. At increasing wavelengths, its response speeds up as the relative light-matter interaction of the probe with the AZO increases.

In the fabricated TiN-AZO resonator, the 325-nm pump excites electrons in TiN and AZO, red shifting the resonance in the visible[60], and blue-shifting the resonance in the near-infrared (NIR) range[12] (**Fig. 1c**). Reflectance modulation occurs at nanosecond timescales for visible wavelengths and speeds up as the wavelength increases, until it reaches picosecond timescales at NIR wavelengths (**Fig. 1d**).

This transition happens because, in the visible regime, the probe light is mostly localized in and interacts strongly with TiN, thus the slow carrier dynamics of TiN dominate the dynamics of the switch. Similarly, near the ENZ point of AZO at near-infrared wavelengths, AZO dynamics drives the switching speed. The switching speed between the two material resonances lies somewhere between the TiN and AZO relaxation times. The speed increases as the probe wavelength gets closer to the AZO resonance. Thus, by exploiting the nonlinearities of the two materials, our device can operate at speeds ranging from the GHz to the THz regime. This is the first time the same device has been demonstrated to operate at speeds differing by two orders of magnitude with the same optical pump.

The next sections describe the design and characterization of the hybrid device. We start with the steady-state reflectance characterization to determine the nature of the observed resonant dips.

## 4. Epsilon-Near-Zero Enhanced All-Optical Switch Design and Fabrication

A typical Ferrell-Berreman resonator, which comprises an epsilon-near-zero medium with a subwavelength thickness on a reflective backplane[35,63], is engineered to directly couple free-space *p*-polarized light into the medium (Fig. 2a). We perform reflectance measurements on our TiN-AZO double-layer at the 50° angle of incidence. The device shows two dips near the TiN and AZO ENZ points in the reflectance spectrum of *p*-polarized light (**Fig. 2b**). Around 1360 nm *p*-polarized light near the AZO ENZ couples into a radiative Berreman mode, resulting in a reflectance dip. TiN is metallic near the ENZ of AZO, serving as the back reflector required for the Berreman mode. On the other hand, near the ENZ of TiN, AZO is a dielectric allowing light to pass into the TiN. The high refractive index of silicon allows it to act as a reflective backreflector, allowing light to couple into the radiative ENZ mode[63] in the TiN, resulting in the reflectance dip around 485 nm. Thus, our device supports a radiative ENZ mode in the visible and a Ferrell-Berreman mode in the infrared. The *s*-polarized light spectra have multiple Fabry-Perot resonance dips throughout the wavelength range measured. Spectroscopic ellipsometry followed by fitting with a Drude-Lorentz model shows that TiN has an epsilon-near-zero point at 485 nm, and AZO at 1360 nm (**Fig 2c,d**).

The planar structure of this dual-resonant device also ensures that additional nanostructuring-induced recombination channels[13,64] do not affect its switching response.

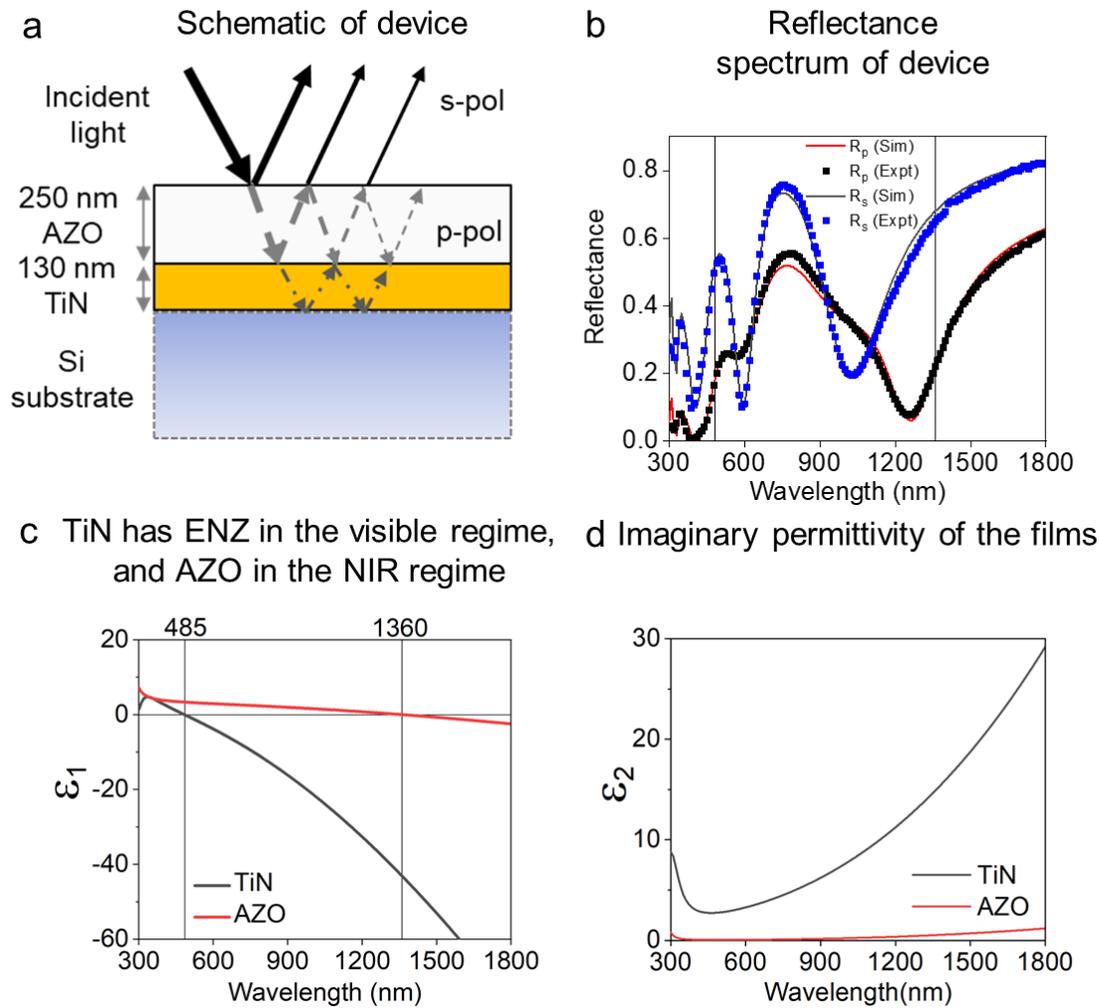

Figure 2 (a) The double-layer TiN-AZO resonator cross-section. (b) Reflectance spectra for the s- and p–polarized light with an incidence angle of 50 degrees. Near the ENZ of the respective films, p-polarized light couples into either the radiative ENZ or FB mode. S-polarized light can couple into Fabry-Perot modes or is reflected. (c) The real part of the permittivity of the as-deposited TiN and AZO films; the vertical lines show the ENZ points of the films. (d) The imaginary permittivity of the films.

## 5. Pump-probe spectroscopy of individual films

To understand the effect of each film's temporal response on the overall behavior of the device, we pre-characterized the TiN and AZO films, individually grown on silicon, close to their respective ENZ points. **Figure 3a** shows the experimental setup used to measure the response of TiN on silicon. An 800 nm, 2 kHz amplified Ti: sapphire laser is split into two branches a pump branch and a probe branch. The pump branch is directed into an OPA (Topas) and converted to 325 nm. The probe branch is delayed using an electronically controlled delay stage and then focused into a Beta Barium Borate (BBO) crystal to generate supercontinuum white light. The pump and probe beams are spatially overlapped on the sample. The pump-fluence was fixed at 1.5 mJ/cm$^2$/pulse.

The temporal response of charge carriers in TiN has been studied in detail by Kinsey et al.[16] and Diroll et al.[61] Upon excitation by an optical pump, the reflectance spectrum of the TiN redshifts. The redshift results in broadband reflectance modulation, with a positive reflectance change to the left of the minimum and a negative change to the right. In our pump-probe measurements, we see a similar TiN spectral response (**Fig. 3b**). **Supporting information S4** has more details on the dynamic characterization of TiN films. The response can be accurately fitted with a two-time-constant model (Eq. S2), with the dynamics attributed to lattice cooling[61]. The overall response time of our TiN film is more than 1 nanosecond (**Fig. 3c**), which determines the relatively slow switching speeds of TiN mirrors.

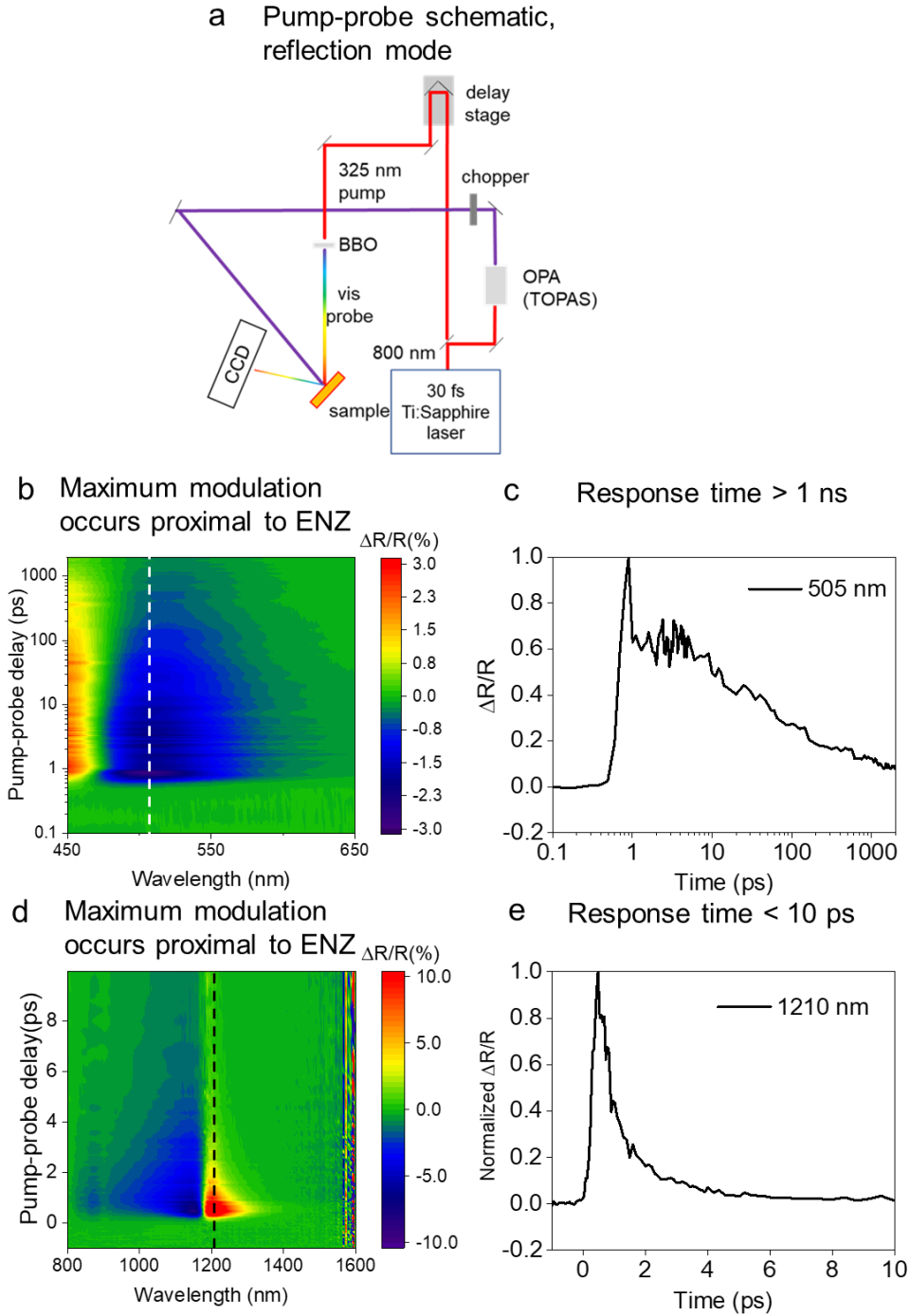

Figure 3 (a) The pump-probe spectroscopy setup in reflectance mode with a 325 nm pump and visible probe. (b) Transient reflectance modulation versus time for TiN on Si probed at visible wavelengths. The response along the vertical dashed line near the ENZ wavelength of 505 nanometers is plotted in c. (c) The modulation has a zero-to-zero response time of several nanoseconds at 505nm (d) Transient reflectance modulation versus time for 500 nm AZO on silicon probed at near-infrared wavelengths. The response along the vertical dashed line at the near ENZ wavelength of 1210 nm is plotted in e. (e) The modulation has a zero-to-zero response time of less than 10 ps at 1210 nm.

Next, we studied the temporal response of AZO films under the same pump. We deposited a 300-nm-thick AZO film on silicon for these studies (See Supporting Information for the growth and modeling details). We performed the pump-probe spectroscopy on the AZO using the setup shown in **Fig. 3a,** with a sapphire replacing the BBO crystal to produce the supercontinuum NIR probe (800-1600 nm). The pump was formed using the same method described in the previous section.

The interband pump generates free carriers in AZO, increasing the absorption and decreasing the refractive index by modifying a Drude dispersion term[12]. The reflectance spectrum blueshifts, decreasing the reflectance to the left of the minima and increasing to the right (**Fig. 3d**). As electrons recombine, the sample returns to its original refractive index. **Supporting information S5** has more details on the dynamic characterization of AZO films. The transient reflectance can be fitted with a single sub-picosecond time constant (Eq. S3). This result points to a dominant, single recombination channel. In prior work, it was identified as the defect-assisted Shockley-Read-Hall mechanism[12]. The overall relaxation time shows little change if the pump fluence is varied. The reflectance modulation falls to zero in less than 10 ps (**Fig. 3e**), allowing for the very fast switching of AZO switches. We note that the substrate on which AZO is grown may affect the defect density or the switching response. To account for this, we also grew a thick AZO film on fused silica and observed similar picosecond scale switching speeds (**Supporting Information S5**).

## 6. Metasurface pump-probe spectroscopy

After pre-characterizing the individual TiN and AZO films' dynamic response, we performed pump-probe measurements on the device in the reflection mode. The pump-probe setup is the same in **Fig. 3a** used for the individual film characterization. We used both visible and NIR light to probe the device. The pump fluence was kept at 1.5 mJ/cm$^2$, the same as in the previous section. The maximum reflectance modulation at visible and near-infrared wavelengths occurs proximal to the steady state-reflectance dips previously measured for individual TiN and AZO films. In particular, these dips occur near the ENZ points of the TiN and AZO films.

At visible wavelengths, there is an overall decrease in the reflectance above the minimum modulation wavelength and an increase below this wavelength (**Fig. 4a**) with a nanosecond response. The amplitude and the response dynamics are similar to the photoexcited TiN film in the previous section (**Fig. 3b,c**). The overall response time of the modulation is slow (ns range).

In the near-infrared regime, the dynamic spectral change is similar to that of AZO. The maximum change is negative below and positive above the modulation minimum wavelength (**Fig. 4b**). This corresponds to a transient blueshift in the reflectance dip with a picosecond response, similar to that of the AZO film (**Fig. 3d,e**). The relaxation time of the response is under 10 picoseconds.

Finite Element Method (FEM) simulations (COMSOL Multiphysics) with our material's measured optical properties show that at 325nm (under normal incidence), 48% of the pump light is absorbed in the TiN layer and 39% in the AZO layer. Thus, upon excitation by the pump, electrons in both AZO and TiN are excited, modulating the permittivity of both materials. Carriers in AZO relax at ultrafast timescales while the permittivity of TiN takes longer to settle to steady-state due to its slower lattice cooling relaxation process.

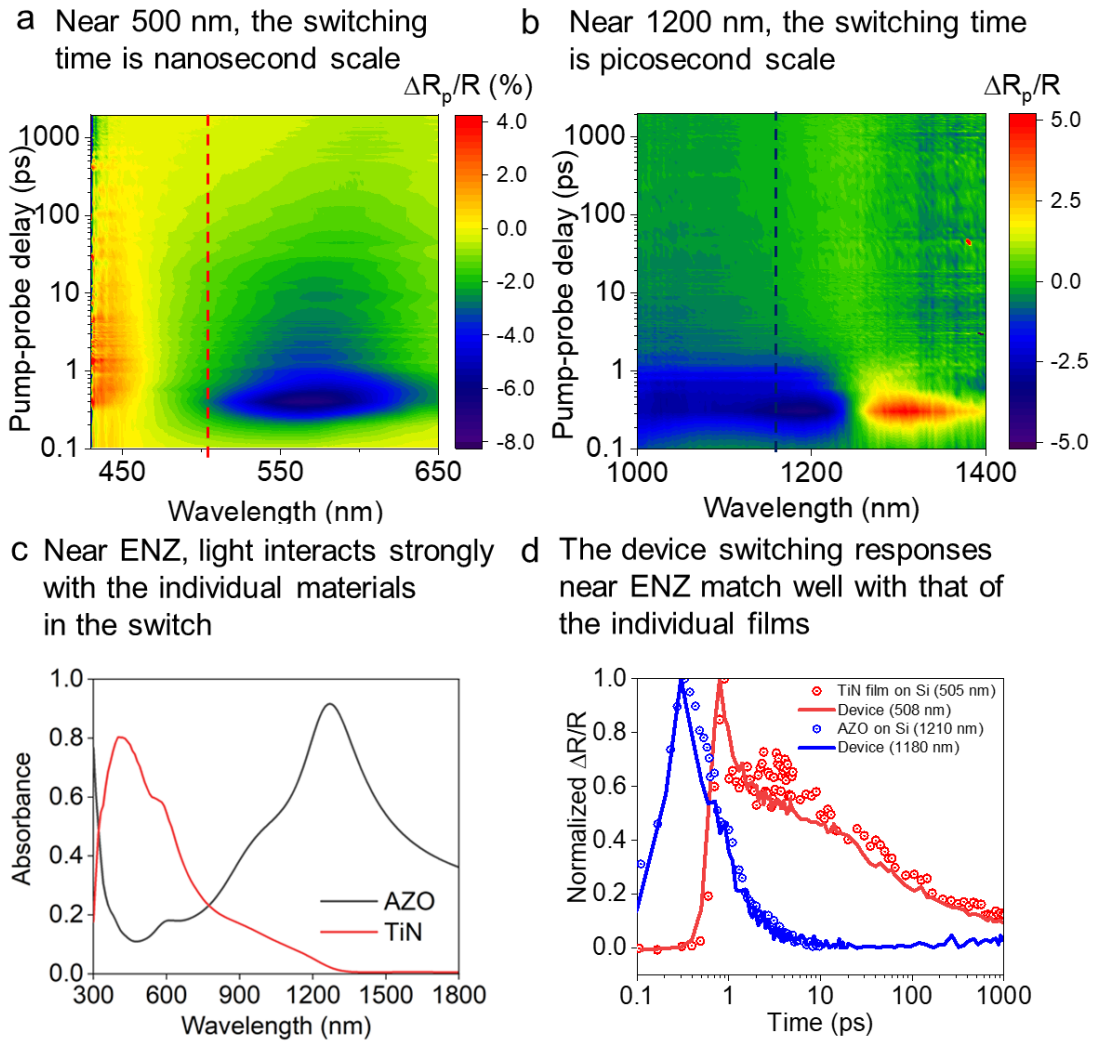

Figure 4 (a) Transient reflectance modulation versus time for the absorber with visible probes and (b) near-infrared probes. The red and blue vertical dashed lines correspond to the device dynamics at 508 nm and 1180 nm, respectively. These are plotted in Fig. 4d as solid lines. (c) The absorbance of the p-polarized light in the device at an angle of 50° as calculated by the Finite Element Method, showing that light interacts strongly with individual films near their respective ENZ points. (d) The transient modulation response of the device at a probe wavelength of 508 nm (solid red line) and 1180 nm (solid blue line). The points are the modulation responses of the TiN film at 505nm (red points) and the AZO film at 1210nm (blue points). The slower TiN dynamics are shifted by 0.4 ps for clarity.

FEM simulations (COMSOL) using the extracted optical properties of TiN and AZO show that 80% of p-polarized light is absorbed in the TiN near its ENZ point in the visible. Similarly, 90% of p-polarized is absorbed by the AZO near its ENZ point in the NIR (**Fig. 4c**). Thus the visible probes interact strongly with the TiN layer near its ENZ, resulting in slower, nanosecond switching times. On the other hand, near its ENZ the near-infrared probe interacts strongly with the AZO film, resulting in ultrafast, picosecond scale switching time.

To further illustrate the phenomenon, in **Fig. 4d** we overlap the temporal response of the device and that of the individual thick films from the previous section at different wavelengths. At 508 nm, the modulation dynamics of the bilayer device (solid red line) closely follows that of the TiN film on silicon (red dots), as the probe strongly interacts with the titanium nitride layer (**Fig. 4c**). On the other hand, at 1180 nm, the probe strongly interacts with the AZO layer, as shown by the strong absorbance of *p*-polarized light in the AZO. As a result, the speed of the modulation at 1180 nm (solid blue line) mirrors that of AZO on silicon (blue dots).

## 7. Effective temporal response of the device

At wavelength regimes where the probe interacts comparably with TiN and AZO, the response has attributes of both materials. **Figure 5a** shows how the power dissipation in individual films varies with increasing wavelength. As we transition from visible to NIR wavelengths, dissipation decreases in the TiN and increases in the AZO. This redistribution demonstrates increasing light-matter interaction with AZO the faster of the two materials.

a

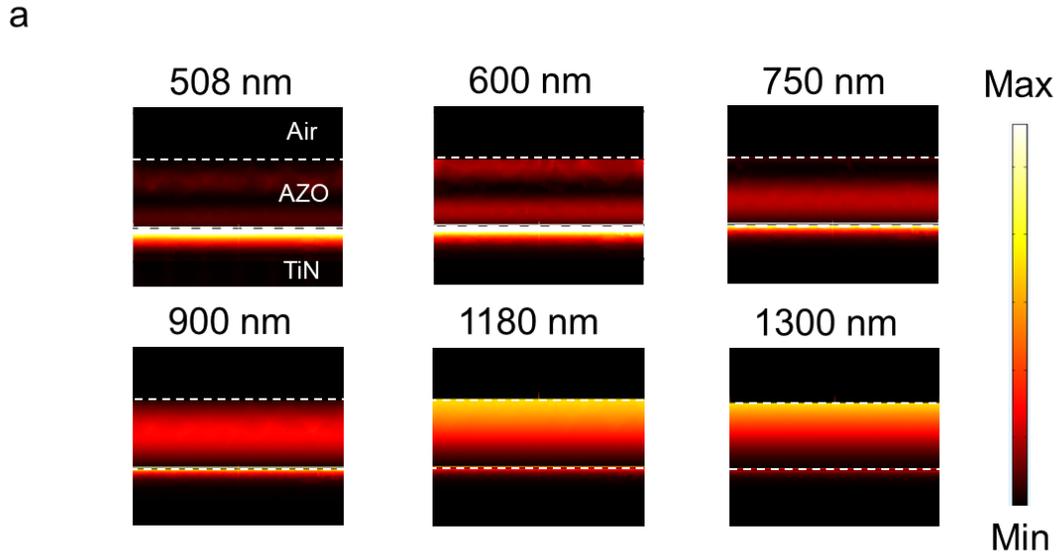

b The switching time gradually decreases as less light interacts with TiN

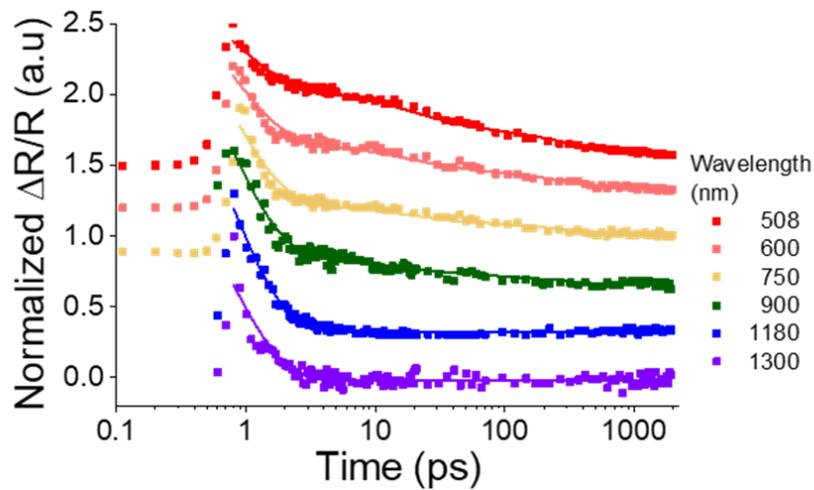

Figure 5 (a) Simulated power dissipation (normalized) in the films computed using measured optical properties of the films. The white dashed lines denote the Air-AZO and AZO-TiN boundaries. As the wavelength increases, more light interacts with the AZO film compared to TiN (b) Experimentally obtained reflectance modulation versus pump-probe delay time. The modulation for each wavelength is normalized to the maximum response at each wavelength and shifted to the same time for easier comparison. Starting from 1300 nm (purple), the datasets for each wavelength are vertically shifted by 0.3 from the next (from purple to red) for easier comparison. The points represent experimental results, and the solid lines represent fitted values. As the probe wavelength increases, the relaxation rate transitions gradually from nanosecond to picosecond-scales

Plotting the temporal response of the device at different probe wavelengths shows a gradual decrease in the overall relaxation time with increasing wavelength (**Fig. 5b**). The effective temporal response of the device can be modeled as a weighted sum of the individual responses of the materials:

$$\left(\frac{\Delta R}{R}\right)(t) = \alpha\left(Ae^{\frac{-t}{\tau 1}} + Be^{\frac{-t}{\tau 2}}\right) + \beta De^{\frac{-t}{\tau}} + \Gamma$$

where $\alpha$ is the weighted contribution of TiN (with the 2-time constants $\tau 1$ & $\tau 2$) and $\boldsymbol{\beta}$ is the weighted contribution of the AZO (single time constant $\tau$). $\boldsymbol{\Gamma}$ is an offset attributed to slower thermal effects and noise due to probe fluctuations. The ratio of $\alpha/(\alpha + \beta)$, the ratio of the absorption in TiN to the total absorption, decreases as expected with increasing wavelength as light (**Supporting Fig. S3b**). Intuitively, as the probe wavelength increases, more light interacts with the AZO versus TiN. Thus, AZO dictates the modulation dynamics, resulting in faster response dynamics. Therefore, it is possible to control the switching speed of the same device from nanosecond to picosecond scale by simply changing the wavelength of operation.

## 8. Discussion and Conclusion

While extensive strides have been conducted in the dynamic control of the phase, amplitude, and polarization of light with optical devices, there exists a gap in engineering the temporal dynamics of tunable systems. In this work, by employing an all-optical double-layer switch comprising both fast and slow-responding materials, we demonstrated advanced control over the temporal dynamics of such a hybrid all-optical switch. The device supports two p-polarized light resonances – a radiative visible ENZ mode in the TiN layer and an infrared Ferrell-Berreman mode in the AZO layer. When excited by the same 325 nm wavelength pump, close to these resonances, the device has a switching speed under 10 ps when probed at 1180 nm and over 2 ns when probed at 508 nm. The probe follows the relaxation dynamics of the constituent material near its epsilon near zero point, where the light-matter interaction is the strongest. By sweeping the probe between these

ENZ-related resonances, one can operate the same device at switching speeds spanning two orders of magnitude, enabling a great degree of control over the switching speed.

We show that fast permittivity modulation can be obtained even when materials with relatively slow dynamics are used as a back reflector, to enhance the modulation at the desired operational wavelength. Our findings underline the importance of conducting broadband pump-probe characterization of the full device stack, including spacers, back reflectors, and substrates in addition to the active layer when designing all-optical switches. Furthermore, when designing complex time-varying devices utilizing multiple materials, a good understanding of the dynamic response of each constituent material is required to achieve the desired performance.

We show that the overall temporal response of a dual-material all-optical switch can be represented by a weighted sum of the relaxation behaviors of each of the constituent materials. The zero-to-zero response of an all-optical switch at a particular wavelength is dictated by the material with the stronger light-matter interaction, in other words, by the material that confines the most incident light within it. By changing the probe-light wavelength to interact with different materials in the device, the range of operating speeds and thus bandwidth can be varied by two orders of magnitude. We experimentally demonstrated that controlling the probe-light interaction with the individual TiN or AZO layers enables the large variation of the observed speed. Moreover, we mapped the light-matter interaction of the probe with the materials and correlate their relative absorbances with particular relaxation times, allowing for piecewise design and modeling.

Our approach can be used to "mix and match" materials with significantly different temporal dynamics to achieve multiple switching speeds and bandwidths within the same platform for multiband data transmission. In the field of all-optical switching beyond Moore's Law, optical switches have been demonstrated to operate at femtosecond speeds, offering terahertz or higher switching rates[48]. However, in most practical applications, such switches would still need to communicate with electrical components, whose speed is limited to gigahertz scales. To mitigate the modulation strength versus bandwidth compromise, it is desirable to have the response speed of the modulating effect be only a couple of times faster than the operating speed of the device one is trying to get[65,66]. Thus, the option of tunable switching speeds that match slowly increasing electronic switching speeds offers a way to bridge the gap between electronic and optical communications.

Robust, laser-tolerant materials can be also employed as back reflectors in optical switching schemes and nonlinear optical experiments. Their slower speeds do not affect the ultrafast response of the active medium. Examples of such systems include nonlinear optical phenomena such as frequency shifts[67], negative refraction[68], time refraction[69], and photonic time crystals[70] that require ultrafast changes in the material permittivity at fast, femtosecond timescales, while operating at large laser powers.

Finally, our method can be used to extract the temporal dynamics of individual materials in complex, multilayer structures by controlling where the probe light is strongly confined.

Harnessing control over the all-optical switching speeds by developing a holistic approach to studying the temporal dynamics of time-varying devices and incorporating the contribution of each material will be crucial in developing time-varying devices for telecommunications optics and other nonlinear optical applications.

# Supporting Information for All-Optical Switching: Engineering the Temporal Dynamics with *Fast* and *Slow* Materials

## S1. Growth of titanium nitride on silicon

The TiN is grown by DC reactive magnetron sputtering using a titanium target. A 99.5% pure Ti target of 2-inch diameter is used. The distance from the target to the source is kept at 20 cm to ensure uniformity. The chamber is pumped down to $10^{-8}$ T to prevent oxygen contamination and is backfilled with Ar at a pressure of 5 mTorr. The Ti is sputtered for 2 minutes to clean the top surface with a power of 200W. Then, an Ar:$N_2$ mixture of 1:18 ratios is used for the sputtering process. The substrate is heated to a temperature of 800°C and rotated at 5 rpm.

## S2. Growth of aluminum-doped zinc oxide

We grow the AZO on optically thick TiN layers grown on silicon using the method mentioned above. The substrates are heated at 45°C. The chamber is pumped down to a vacuum of $10^{-6}$ Torr. ZnO is deposited on the TiN films by pulsed laser deposition (PLD) with a KrF excimer laser at a pump fluence of 1.5 J/cm$^2$. The chamber is pumped down to $6\times10^{-6}$ T and then backfilled to 3 mTorr with oxygen. The growth rate is 6 nm/min. We also grow AZO on silicon substrates and fused silica using the same growth parameters.

## S3. Steady-state optical properties of titanium nitride and AZO

The TiN films are measured with the variable angle spectroscopic ellipsometry (VASE) at angles of 50 and 70°. We used a Drude Lorentz model to describe the optical properties of the TiN, with one Drude oscillator and two Lorentz oscillators.

$$\varepsilon = \varepsilon_1 + i\varepsilon_2 = \varepsilon_\infty - \frac{A_0}{(\hbar\omega)^2 + iB_0\hbar\omega} + \sum_k^n \frac{A_k}{E_k^2 - (\hbar\omega)^2 - iB_k\hbar\omega} \quad \text{(Eq. S1)}$$

where $\varepsilon_1$ and $\varepsilon_2$ are the real and imaginary parts of the dielectric permittivity, $\hbar$ is the reduced Plank constant, $\omega$ is the probe angular frequency, and $\varepsilon_\infty$ is an additional offset. The negative (Drude) part shows the contribution of the free carriers to the permittivity: $A_0$ is the square of the plasma frequency (in eV) and $B_0$ is the damping factor of the Drude oscillator. The effect of bound electrons and interband transitions is modeled by a summation of the Lorentz oscillators, with $A_k$, $B_k$, and $E_k$ being the amplitude, broadening, and center energy of individual Lorentz oscillators.

Table S1. Drude-Lorentz parameters for the as-deposited TiN and AZO films

| Thickness | $A_0$ | $B_0$ | $A_1$ | $B_1$ | $E_1$ | $A_2$ | $B_2$ | $E_2$ | $\varepsilon_\infty$ |
|---|---|---|---|---|---|---|---|---|---|
| TiN | 43.3 | 0.196 | 310.6 | 76.2 | 4.61 | 31.6 | 0.993 | 4.15 | 3.64 |
| AZO on TiN | 2.91 | 0.139 | 17.8 | 0.147 | 4.55 | - | - | - | 2.54 |
| AZO on Si | 3.03 | 0.086 | 4.52 | 0.122 | 4.27 | | | | 3.08 |
| AZO on FS | 2.80 | 0.079 | 309 | 50.6 | 26.6 | - | - | - | 2.88 |

## S4. Temporal dynamics of TiN

**Fig. S1** shows the temporal dynamics of TiN when pumped with 325 nm light at a pump fluence of 1.5 mJ/cm². Prior work by Kinsey et al. and Diroll et al.[1,2] show TiN to have an ultrafast electron-phonon relaxation rate that is see when probed near the ENZ point. Our experiment shows similar results, as seen from the sharp initial decay of the probe at 505 nm followed by the slower decay (black curve). Far from the ENZ, the ultrafast component is no longer seen, and the slower components span a nanosecond timescale (green graph). For this study, we focused on the slower dynamics of the TiN films. The overall dynamics is fitted with a two time-constant model, comprising of two exponential components, with time constants of the faster component of ≈20 ps for TiN followed by a slower time constant of ≈250 ps, both attributed to lattice cooling via diffusion and the radiation of heat from the measured sample spot, including through available interfaces.

$$\left(\frac{\Delta R}{R}\right)(t) = A e^{\frac{-t}{\tau_1}} + B e^{\frac{-t}{\tau_2}} + C \qquad (\text{Eq. S2})$$

where $A, B, C, \tau_{1,2}$ are the relaxation fitting constants summarized in Table S2.

Table S2. Fitting parameters of TiN transient reflectance modulation

| Wavelength (nm) | A | B | C | $\tau_1$ (ps) | $\tau_2$ (ps) |
|---|---|---|---|---|---|
| 505 | 0.33 | 0.26 | 0.096 | 17.3 | 287 |
| 750 | 0.29 | 0.28 | 0.24 | 24.3 | 242 |

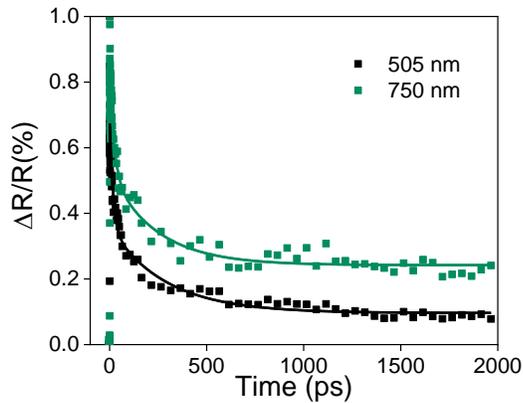

Figure S1. Normalized Transient reflectance of TiN at different wavelengths

## S5. Temporal dynamics of AZO on Si with different powers, and that of AZO on fused silica

We perform pump-probe spectroscopy on the AZO films at various pump fluences. Fig. S2a shows the normalized reflectance spectra of AZO at various pump fluences. The overall decay time is invariant with increased pump fluence. Prior studies attribute this ultrafast response time to defect-assisted Shockley-Read-Hall recombination, a common recombination pathway in heavily doped TCOs [3,4].

The carrier relaxation time of doped oxides may also be dependent on the growth substrate [5]. To study how much the recombination time depends on the substrate, we grew AZO fused silica. Pump probe spectroscopy of the AZO grown on fused silica shows it to have a recombination time under 10 ps.

$$\left(\frac{\Delta R}{R}\right)(t) = D e^{\frac{-t}{\tau}} + E \qquad (\text{Eq. S3})$$

where $D, E, \tau$ are the relaxation fitting constants summarized in Table S3.

Table S3 Fitting parameters of AZO transient reflectance modulation at a wavelength of 1.2 µm

| Pump Fluence (mJ/cm$^2$) | D | E | $\tau$ (ps) |
| --- | --- | --- | --- |
| 0.18 | 1.76 | 0.056 | 0.715 |
| 0.49 | 1.69 | 0.046 | 0.873 |
| 1.53 | 1.98 | 0.045 | 0.723 |
| 2.48 | 1.65 | 0.088 | 0.834 |

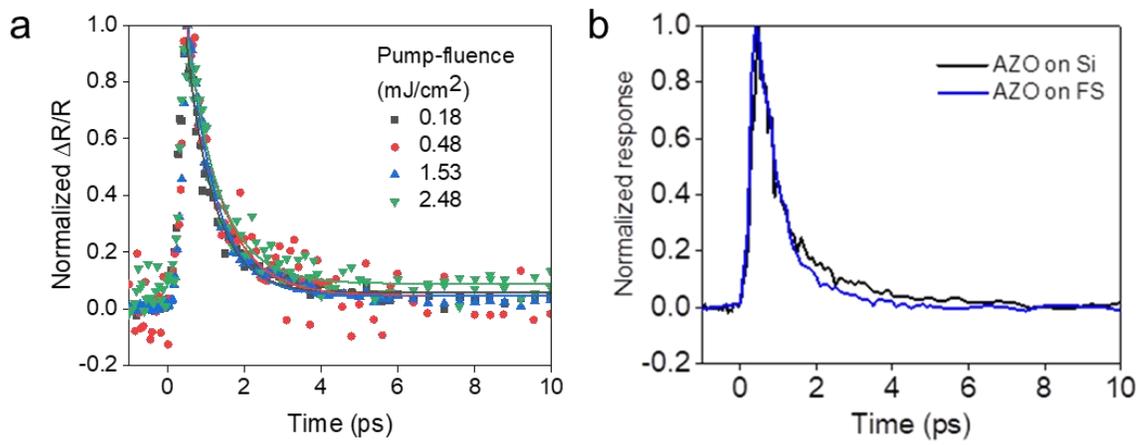

Figure S2. (a) Transient reflectance spectra of AZO on Si at different pump-fluences at 1200 nm (b) Transient reflectance spectrum of AZO on Si overlapped on that of AZO grown on fused silica, showing similar response times.

## S6. Metasurface response as a weighted sum of the material responses

Having extracted the temporal responses of the individual films, the response of the metasurface at a specific wavelength can be modeled as a weighted sum of the temporal response of the TiN and the AZO films.

$$\left(\frac{\Delta R}{R}\right)(t) = \alpha\left(Ae^{\frac{-t}{\tau_1}} + Be^{\frac{-t}{\tau_2}}\right) + \beta D e^{\frac{-t}{\tau}} + \Gamma \qquad \text{(Eq. S4)}$$

$\alpha$ is a weight factor of TiN (with time constants, $\tau_1 = 17.3\ ps$, $\tau_2 = 287\ ps$ and constants $A = 0.33, B = 0.26$, from Table S.2 at 505 nm) and $\beta$ is a weight factor of the AZO (with time constant $\tau = 0.72\ ps$ and constant $D = 1.98$ from Table S.3 at a fluence of 0.18mJ/cm$^2$). $\Gamma$ comprises much slower lattice cooling effects, together with detector noise arising from probe fluctuations. Figure S3 shows the fitted graphs of the experimental data at each wavelength. For these fits, we used the time constants of TiN at 505 nm and of AZO at 1200 nm. Incorporating the wavelength dependent variations of the time constants can be expected to give better fits. Fig. S3b shows the relative weight of the contribution of TiN to the overall response $\alpha/(\alpha + \beta)$ (red curve), overlapped with the simulated absorbance of the TiN (Abs$_{TiN}$) relative to the total absorbance (Abs$_{TiN+AZO}$) at each wavelength. The trends show a strong correlation. The absorbance of probe in a medium is strongly affected by the light-matter interaction between the probe and the medium. This shows that the time constants of the recombination is indeed influenced strongly by the light-matter interaction of the probe with the individual layer.

Table S4 Fitting parameters for the metasurface dynamic response

| Wavelength (nm) | 508 | 600 | 750 | 900 | 1180 | 1300 |
|---|---|---|---|---|---|---|
| $\alpha$ | 0.837 | 0.583 | 0.380 | 0.347 | 0.008 | -0.018 |
| $\beta$ | 0.476 | 0.702 | 0.965 | 1.147 | 1.338 | 1.045 |
| $\Gamma$ | 0.08 | 0.135 | 0.001 | 0.050 | 0.023 | -0.015 |
| Reduced chi-square | 5.04E-4 | 8.27E-4 | 10.7E-4 | 12.4E-4 | 4.36E-4 | 19.3E-4 |

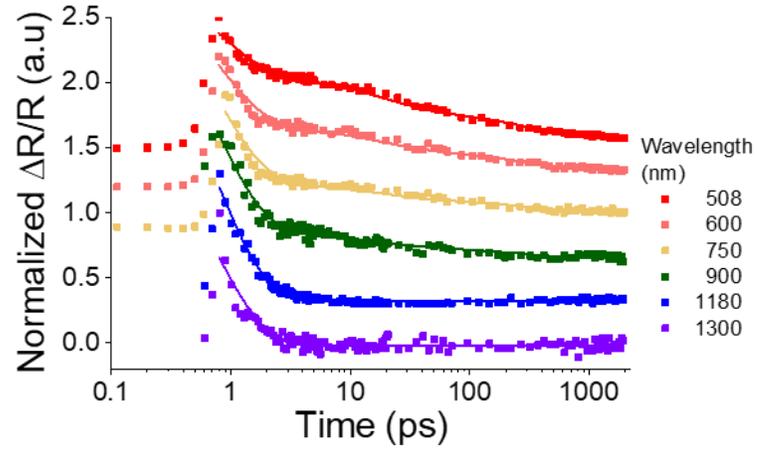

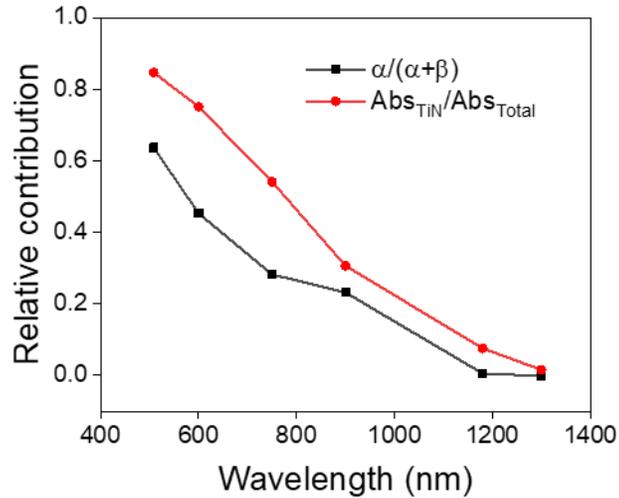

Figure S3 (a) Experimentally obtained temporal response of the metasurface (normalized) at different wavelengths (dots) overlapped with the fits (solid lines). The maximum pump-probe overlap for each experiment was alligned to center at 1 ps, and each curve is vertically translated by 0.3 from the next for clarity. (b) Relative weight of the TiN contribution to the sum of TiN and AZO contributions (red), and the relative absorbance of TiN as a fraction of the total absorbance of the metasurface (black)